\documentclass[prc,twocolumn,showpacs,showkeys,amsmath,amssymb]{revtex4}

\usepackage{graphicx}
\usepackage{dcolumn}
\usepackage{bm}
\usepackage{epstopdf}

\newcommand{\etal}{{\it et al.}, }

\begin{document}

\title{Measuring the Cosmic Ray Muon-Induced Fast Neutron Spectrum\\
by ($\bm n$,$\bm p$) Isotope Production Reactions in Underground Detectors}

\author{Cristiano Galbiati}
\email[Email address: ]{galbiati@princeton.edu}
\affiliation{Department of Physics, Princeton University, Princeton,
NJ 08544, USA} 

\author{John F. Beacom}
\email[Email address: ]{beacom@mps.ohio-state.edu}
\affiliation{Department of Physics, Ohio State University, Columbus,
OH 43210, USA}
\affiliation{Department of Astronomy, Ohio State University, Columbus,
OH 43210, USA
\medskip}

\date{\today}

\begin{abstract}
While cosmic ray muons themselves are relatively easy to veto in
underground detectors, their interactions with nuclei
create more insidious backgrounds via: (i) the decays of long-lived
isotopes produced by muon-induced spallation reactions inside the
detector, (ii) spallation reactions initiated by fast muon-induced
neutrons entering from outside the detector, and (iii) nuclear recoils
initiated by fast muon-induced neutrons entering from outside the
detector.  These backgrounds, which are difficult to veto or shield
against, are very important for solar, reactor, dark matter, and other
underground experiments, especially as increased sensitivity is
pursued.  We used \textsc{fluka} to calculate the production rates and
spectra of all prominent secondaries produced by cosmic ray muons, in
particular focusing on secondary neutrons, due to their importance.
Since the neutron spectrum is steeply falling, the total neutron
production rate is sensitive just to the relatively soft neutrons, and
not to the fast-neutron component.  We show that the neutron spectrum
in the range~$\sim$\,10--100\,MeV can instead be probed by the
$(n,p)$-induced isotope production rates $^{12}$C$(n,p)^{12}$B and
$^{16}$O$(n,p)^{16}$N in oil- and water-based detectors.  The result
for $^{12}$B is in good agreement with the recent KamLAND measurement.
Besides testing the calculation of muon secondaries, these results are
also of practical importance, since $^{12}$B ($T_{1/2} = 20.2$\,ms, $Q
= 13.4$ MeV) and $^{16}$N ($T_{1/2} = 7.13$\,s, $Q = 10.4$ MeV) are
among the dominant spallation backgrounds in these detectors.
\end{abstract}

%




\maketitle


\section{Introduction}
\label{sec:introduction}

To reduce the cosmic-ray muon background, experiments to measure rare
processes must be sited underground, where the muon flux is greatly
attenuated, and surrounded by an active veto system, to tag the
residual muons.  Even with these standard measures, muons are still
responsible for significant backgrounds in underground experiments,
via the secondary particles created by muon interactions with nuclei.
If the muon interacts inside the detector, the secondary shower
particles create unstable isotopes; some have long lifetimes, making
it hard to associate them with particular muons.  If the muon
interacts outside the detector, it cannot be tagged, and ``invisible"
secondaries, especially neutrons, can penetrate the detector
shielding.  These neutrons can then initiate spallation reactions or
nuclear recoils inside the detector.  While these muon-induced
backgrounds are already given serious consideration at present, the
next generation of underground neutrino, dark matter, and double-beta
decay experiments will require both lower backgrounds and a better
quantitative understanding of their characteristics.

Fundamental to understanding these backgrounds is the rate and
spectrum of muon-induced neutrons~\cite{olga}.  The neutron spectrum
is steeply falling over orders of magnitude in neutron energy, but not
uniformly so, indicating complexity in its formation.  The total rate
of neutron production depends primarily on the soft-neutron spectrum,
and has been well measured.  In Table~\ref{tab:muons} we summarize the
main characteristics of the muon flux underground at several relevant
depths~\cite{hagner,ambrosio,cribier,kamland,waltham,tagg,kudryavtsev,pdg}.
With increasing depth, the muon flux falls quickly and the muon
average energy rises at first quickly and then much more slowly.  The
capture rates of neutrons produced by muons are also noted.  The Gran
Sasso rate was measured with the Borexino Counting Test
Facility~\cite{borex,deutsch1,galbiati}.  The rates at other depths
were calculated using the scaling law given by Ref.~\cite{olga}; the
result for Kamioka is fully consistent with the rate of 2940/kton\,day
measured by KamLAND~\cite{kamland2}.

What is needed now is a more quantitative understanding of the neutron
spectrum at moderate and high energies, as emphasized by
Ref.~\cite{wang}.  In this paper, we consider the production of
unstable isotopes as a new and direct probe of the moderate-energy
neutron spectrum.  We focus our attention here on the $(n,p)$
reactions in oil- and water-based detectors, and show that their rates
are a sensitive probe of the $\sim$ 10--100 MeV neutron spectrum,
which we calculate using \textsc{fluka}.  The predicted rate of
$^{12}$C$(n,p)^{12}$B production is in very good agreement with the
rate measured in the KamLAND experiment~\cite{kamland2}.  These {\it
in-situ} measurements are an important complement to measurements made
at accelerators, namely the experiment of Ref.~\cite{hagner} at CERN
using muon beam energies of 100 and 190 GeV.

The ultimate goal of these studies is to characterize the muon-induced
neutron spectrum precisely, at all energies, as well as the yields of
unstable isotopes produced by
muon secondaries.  Here we make a step towards this goal by focusing
on the reactions $^{12}$C$(n,p)^{12}$B and $^{16}$O$(n,p)^{16}$N in
oil- and water-based detectors.  We show below that at depths greater
than a few hundred m.w.e., these are the only significant production
channels for $^{12}$B in oil-based detectors and $^{16}$N in
water-based detectors, respectively.  These are among the most
significant spallation products in these two important types of
detectors.  By considering these single isotopes, with the same masses
as the parents, we can isolate just the $(n,p)$ production channel,
and hence directly probe the muon-induced neutron spectrum.  Even
though the rates of these $(n,p)$ reactions are well below the total
neutron production rates, they are still quite large: about 60/kton\,day
(calculated and measured) for $^{12}$B in KamLAND~\cite{kamland2},
and 50/kton\,day (calculated) for $^{16}$N in Super-Kamiokande~\cite{sk},
both before cuts normally  designed to suppress these
and other spallation products.  Both $^{12}$B ($T_{1/2} = 20.2$\,ms,
$Q = 13.4$ MeV) and $^{16}$N ($T_{1/2} = 7.13$\,s, $Q = 10.4$ MeV) are
unstable to $\beta^-$ decay, and their high production rates and
endpoint energies make them significant backgrounds; the very long
lifetime of $^{16}$N makes it especially pernicious.

\begin{table}[t]
\caption{\label{tab:muons}
Depth, muon flux, muon average energy, and neutron capture rate at
sea level, 500 m.w.e., and the Kamioka, Gran Sasso, and Sudbury
underground laboratories.}
\begin{ruledtabular}
\begin{tabular}{ldcdc}
	&\multicolumn{1}{c}{Depth}
	&\multicolumn{1}{c}{$\Phi_\mu$}
	&\multicolumn{1}{c}{$\langle E_\mu \rangle$}
	&\multicolumn{1}{c}{$p(n,\gamma)d$} \\
	&\multicolumn{1}{c}{[m.w.e.]}
	&\multicolumn{1}{c}{[$\rm \mu/m^2\,h$]}
	&\multicolumn{1}{c}{[GeV]}
	&\multicolumn{1}{c}{[events/kton day]} \\
\hline
Sea Level		&0		&6.0$\times10^5$	&4		&7.2$\times10^6$ \\
500\,m.w.e.	&500	&610			&100	&8.0$\times10^4$ \\
Kamioka		&2700	&9.6				&285	&3000 \\
Gran Sasso	&3800	&1.2				&320 	&400 \\
SNOLab		&6000	&0.012			&350 	&4.3 \\
\end{tabular}
\end{ruledtabular}
\end{table}

Section~\ref{sec:secondaries} describes our calculation of the
production rate of the secondaries in muon showers.
Section~\ref{sec:12b} offers a precise {\it ab initio} calculation of
the $^{12}$B production rate at different depths, and a direct
comparison of our results with the measured production rate at the
Kamioka depth measured by KamLAND.  Section~\ref{sec:16n} offers a
similar calculation for the production rate of $^{16}$N in water.  We
draw our conclusions in section~\ref{sec:conclusions}.


\section{Muon Production of Secondaries}
\label{sec:secondaries}

\begin{figure}[t]
\includegraphics[width=3.25in]{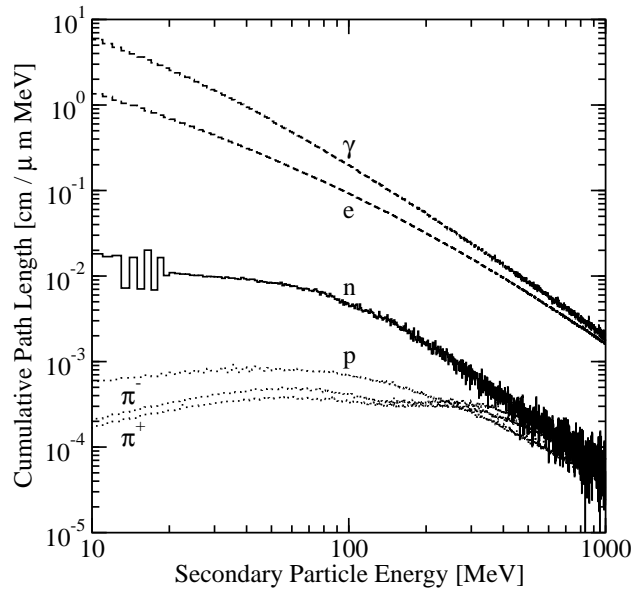}
\caption{\label{fig:secondaries_path} Cumulative path length $dL(E)/dE$ of
secondaries (in cm of path length per meter of $\mu$ track, and in
secondary particle energy bins of 1\,MeV) generated by muons at
285\,GeV, appropriate to the depth of Kamioka.}
\end{figure}

Using the known muon flux underground, we used
\textsc{fluka}~\cite{fluka} to calculate the production rates,
energies, and path lengths of all prominent secondaries, i.e.,
$\gamma$~rays, electrons (and positrons), neutrons, protons, and $\pi$
mesons.  The \textsc{fluka} program is a Monte Carlo code able to
simulate particle showers by propagating particles according to
standard interactions.  The \textsc{fluka} code has been validated for
its use in muon-induced showers in a number of studies.  Most notably,
Wang {\it et al.}~\cite{wang} used \textsc{fluka} to reproduce
experimental results of the production rate of neutrons by muons in
liquid scintillator at several depths, and Kudryavtsev {\it et
al.}~\cite{kudryavtsev} performed extensive studies on the energy
spectrum and range of neutrons produced underground in cosmic-ray
induced showers.

Our \textsc{fluka}-based code was developed in the context of a study
of the production rate of $^{11}$C cosmogenic isotopes in oil-based
detectors underground~\cite{galbiati}.

We simulated showers originating from muons at several relevant
energies: 100\,GeV (corresponding to the beam experiment of
Ref.~\cite{hagner}, and also the average muon energy at a depth of
500\,m.w.e.), 285\,GeV (the average energy at Kamioka), 320\,GeV (the
average energy at Gran Sasso), and 350\,GeV (the average energy
at Sudbury).  The use of the average muon energy should be
adequate given that the cross sections for muon-induced processes scale
nearly like the energy~\cite{olga}.  Only $\mu^-$ were simulated, though the results (except for
muon capture) would be very similar for $\mu^+$.  The target material
in the simulation was the solvent of the liquid scintillator for
Borexino, trimethylbenzene (C$_9$H$_{12}$), with density
0.88\,g/cm$^3$ (incidentally, this makes up 20\% of the solvent used
in KamLAND~\cite{kamland}).  The results should not vary greatly with
other organic solvents, given that typical values of the density and
mass ratio between carbon and hydrogen are close to those of
trimethylbenzene.  Additionally, due to the similar relevant
properties, the results for water should also be similar.  We tracked
muons for 100\,meters, and for each of the prominent secondaries
calculated the cumulative path of the particles as a function of the
particle energy, with a 10\,GeV upper cutoff.

\begin{figure}[t]
\includegraphics[width=3.25in]{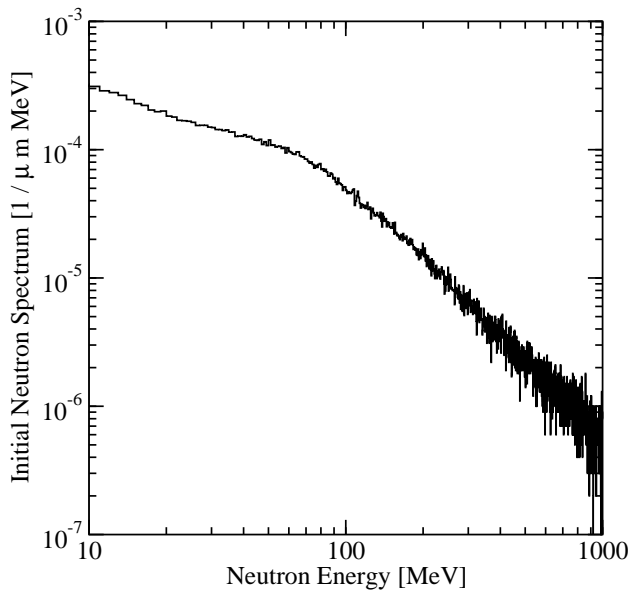}
\caption{\label{fig:neutrons_spectrum} Energy spectrum of neutrons
(per meter of $\mu$ track, and in neutron energy bins of 1\,MeV)
produced by muons at 285\,GeV, corresponding to
Fig.~\ref{fig:secondaries_path}.}
\end{figure}

As a representative example, in Fig.~\ref{fig:secondaries_path} we
show results for secondaries below 1\,GeV produced by muons at
320\,GeV.  For each particle, this figure shows the cumulative path
length $dL(E)/dE$ traveled by all particles of that type at each 1\,MeV bin of
energy.  The relative heights reflect both the particle multiplicities
and also how much path length length they accumulate at each energy
(and hence on the mechanisms of energy and particle loss).  The
calculation includes all real secondary particles in the shower,
including the abundant flux of bremsstrahlung photons from the muons.
It is worthwhile to note that the usually-defined ``range" of the
secondary particles is not directly related to the cumulative path
length as reported in Fig.~\ref{fig:secondaries_path}: in fact, the
trajectory of each secondary particle is broken in a large number of
track segments, each one of them corresponding to the energy of the
particle in that track segment; thus each secondary particle
contributes to a large number of bins in the plot, from the initial
energy down to lower energies as the particle gets slowed down along
its track.

As noted, the neutron secondaries are of special practical importance,
and here we focus just on them and the isotopes they produce by
$(n,p)$ reactions.  Future studies which consider other produced
isotopes will need to consider other secondaries too.  In
Fig.~\ref{fig:neutrons_spectrum}, we show the energy spectrum of the
neutron secondaries.  In this range, the neutron spectrum calculated
here can be described as a power law $\sim E^{-0.5}$
over~$\sim$\,10--100\,MeV, and a power law $\sim E^{-2}$
over~$\sim$\,100--1000\,MeV.  Our results are consistent with those of
Ref.~\cite{kudryavtsev}, which are also based on a \textsc{fluka}
calculation.


\section{$^{12}$B Production in Oil}
\label{sec:12b}


\subsection{Production Reaction ($\bm \mu^-$,$\bm \nu_\mu$)}

While at sea level, the capture of stopped $\mu^-$ on $^{12}$C is the
dominant means of producing $^{12}$B, this is no longer true more than
a few hundred meters underground, due to the steeply falling fraction
of stopping muons.

The rate of stopping muons as a function of depths was inferred from
the muon flux reported in Table~\ref{tab:muons} and from the ratio of
stopping to throughgoing muons from Ref.~\cite{cassiday}.  At the
Kamioka depth the expected rate of stopping muons is about
365/kton\,day (this is consistent with the Super-Kamiokande
measurement of 220/kton\,day, after taking into account the detection
efficiency of 0.65~\cite{blaufuss}).  Only negative muons can undergo
nuclear capture, and the fraction of negative muons is
44\%~\cite{blaufuss}.  The fraction of negative muons undergoing
capture on $^{12}$C in hydrocarbons is 7.7\%~\cite{suzuki,measday}.
For muons undergoing capture, the branching ratio in the channels
resulting in production of a bound $^{12}$B state is
18.6\%~\cite{measday}.  Thus the expected production rate of $^{12}$B
in KamLAND amounts to 2.3/kton\,day.

The expected rate of $^{12}$B production by $\mu^-$ capture at other
depths was also calculated similarly, and the results are summarized
in Table~\ref{tab:12b-rates}.  It is important to bear in mind that
beyond about 500\,m.w.e., the muon average energy and hence all
secondary production rates quoted per meter of muon track, vary only
slowly with depth.  Accordingly, the focus of this paper is the
relative rates of different secondary interactions.

At shallow depths, where the $^{12}$C$(\mu^-,\nu_\mu)$ channel is
important, its rate can be very large.  For example, the $^{12}$B rate
is about 11 Hz in the inner 0.680 kton of MiniBooNE (at sea level),
where it is a significant background for the supernova detection
trigger~\cite{sharp}.


\subsection{Production Reaction ($\bm n$,$\bm p$)}

To evaluate the contribution from ($n$,$p$) reactions, we used a
technique originally developed to calculate the production rate of the
$^{11}$C isotope in organic liquid scintillators~\cite{pep}, and
recently exploited to calculate the production rate of cosmogenic
isotopes in xenon detectors~\cite{bender}.  As shown in
Fig.~\ref{fig:secondaries_path}, one of the key results obtained with
the \textsc{fluka} calculation is the cumulative path length traveled
by secondaries of each type and energy.  Using this, the effects of
reactions of the secondaries can be calculated easily, without
modifying their transport in \textsc{fluka}, provided that the
reactions considered are much less important than the dominant
particle stopping reactions.  For example, for~$\sim$\,10--100\,MeV
neutrons, the ($n$,$p$) cross sections considered here are
$\sim$\,10\,mb, much smaller than the total nuclear cross sections of
$\sim$\,1\,b.

\begin{figure}[t]
\includegraphics[width=3.25in]{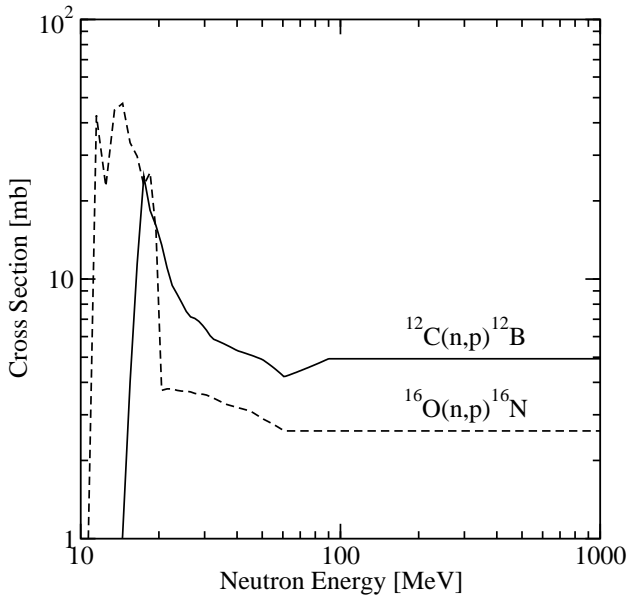}
\caption{\label{fig:cross_sections} Cross sections for
$^{12}$C$(n,p)^{12}$B and $^{16}$O$(n,p)^{16}$N as a function of the
neutron energy in the lab frame.  Data were available up to about
90\,MeV for $^{12}$C and 60\,MeV for $^{16}$O, beyond which we assumed
the cross sections remain constant.}
\end{figure}

For a secondary particle of energy $E$, denote the isotope production
cross section by $\sigma(E)$, and the appropriate target density by
$n$, so that the mean free path is $\lambda(E) = [n \sigma(E)]^{-1}$.
Thus given the cumulative path length $dL(E)/dE$ calculated with
\textsc{fluka}, the expected number of interactions of this type at an
energy $E$ (and per energy range $dE$) 
is simply $[dL(E)/dE]/\lambda(E)$.  It is important to emphasize
that the quantity $dL(E)/dE$ is {\it not} the distance traveled by a secondary
of initial energy $E$; in that case, the dominant stopping reactions
would slow the secondary and reduce its interaction rate.  Instead,
$dL(E)/dE$ is the total amount of path length accumulated by all secondaries
of this type, while they were at the energy $E$.

We will indicate with $R_T$ the total expected number of interactions,
and hence the isotope production rate, given in units of per muon
track length.  The probability for each secondary to have an
interaction in one of the channels of interest and to produce the
cosmogenic isotopes under study here is much smaller than unity, given
that the cross sections for the processes of interest are negligible
with respect to the cross sections of the dominant particle stopping
reactions.  Therefore we can make the following approximation:
\begin{equation}
R_T \simeq \int dE \, \frac{dL(E)}{dE} \, [n \sigma(E)]\,.
\end{equation}
Note that the initial secondary particle energy spectrum is not used
here directly, but only as an input to the second step of the
\textsc{fluka} calculation, which handles all of the particle stopping
reactions after having generated the secondaries.  The relative
weighting of the integral is most conveniently displayed with the
energy on a logarithmic scale, i.e., in terms of $d\log E \sim dE/E$, the
shape of this integrand:
\begin{equation}
R_T \sim \int d\log E \, \left[E \, \frac{dL(E)}{dE} \, \sigma(E)\right]\,.
\end{equation}
In this paper, we consider just the $(n,p)$ reactions for secondary
neutrons.  However, for any secondary particle, any reaction which can
be considered as a perturbation to the main particle stopping
reactions could be treated very similarly.

Returning to the particular case of $^{12}$C$(n,p)^{12}$B, the cross
section was compiled from a number of references~\cite{12b} and is
shown in Fig.~\ref{fig:cross_sections}.  The same figure also includes
the cross section for the process $^{16}$O$(n,p)^{16}$N, also compiled
from a number of references~\cite{16n}.

\begin{figure}[t]
\includegraphics[width=3.25in]{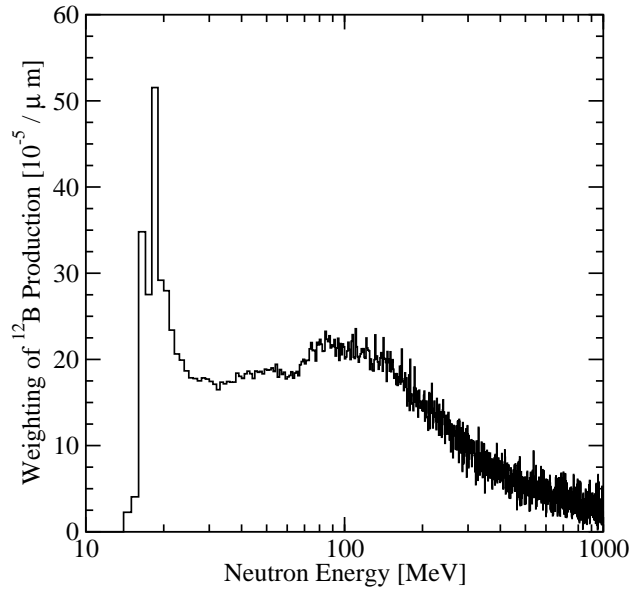}
\caption{\label{fig:special_convolution} Relative weighting of the
$^{12}$C$(n,p)^{12}$B production, considered as an integral in $d \log
E$ (though the data points are evaluated in linear steps of 1 MeV),
for neutrons generated in showers induced by muons at 285 GeV.}
\end{figure}

As noted, a full simulation of muon-induced showers was performed with
\textsc{fluka}~\cite{fluka}, leading to
Fig.~\ref{fig:secondaries_path}.  The product of these two figures and the
energy $E$ (i.e., considered as an integral in $d \log E$) is shown in
Fig.~\ref{fig:special_convolution}, showing that for this reaction,
the most important neutron energies are~$\sim$\,10--100\,MeV, probing
a crucial region of the neutron spectrum shown in
Fig.~\ref{fig:neutrons_spectrum}.  For each decade or fraction thereof
in neutron energy, the relative contribution to the integral can be
immediately estimated by the relative height of the displayed curve.
Due to the $(n,p)$ cross section threshold, this reaction is
insensitive to the very numerous soft neutrons.


\subsection{Other Production Reactions}

We also examined the production channels triggered by $\pi^-$
interactions: $^{12}$B can be produced either by $\pi^-$ capture or by
$\pi^+$ photoproduction, $^{12}$C($\gamma$,$\pi^+$)$^{12}$B.  The
number of $\pi^-$ produced in muon-induced showers at the Kamioka
depth is 4.4$\times$10$^{-3}/\mu$\,m.  The fraction of stopping pions
producing $^{12}$B isotopes in carbon is
9.7$\times$10$^{-4}$~\cite{bistirlich}.  The rate of $^{12}$B
production through $\pi^-$ capture is less than
4$\times$10$^{-6}$/$\mu$\,m and therefore negligible with respect to
the two main channels.  Concerning the
$^{12}$C($\gamma$,$\pi^+$)$^{12}$B exchange reaction, the cross
section is $\sim 1 \mu$b above a threshold of 155\,MeV~\cite{milder},
and thus the yield through this channel is negligible.

Production of $^{12}$B in organic liquid scintillators can also happen
by interaction on the target $^{13}$C.  The low natural isotopic
abundance of $^{13}$C (1.1\%~\cite{isotopes}) and the rate of the
cosmogenic reaction $^{12}$C$\rightarrow ^{11}$C~\cite{pep}, also
resulting in the net loss of a nucleon from the original isotope,
suggest that the production rates through these channels are
negligible.


\subsection{Total Rates for $^{12}$B Production}

\begin{table}[t]
\caption{\label{tab:12b-rates} Production rates for $^{12}$B in muon
induced showers at different depths $D$, given both per muon track
length, and per volume and time.  The experimental number reported by
KamLAND~\cite{kamland2} is also noted for comparison.}
\begin{ruledtabular}
\begin{tabular}{lddddd}
$D$ [m.w.e.]
	&\multicolumn{1}{c}{0}
	&\multicolumn{1}{c}{500}
	&\multicolumn{1}{c}{2700}
	&\multicolumn{1}{c}{3800}
	&\multicolumn{1}{c}{6000} \\
$\langle E_\mu \rangle$ [GeV]
	&\multicolumn{1}{c}{4}
	&\multicolumn{1}{c}{100}
	&\multicolumn{1}{c}{285}
	&\multicolumn{1}{c}{320}
	&\multicolumn{1}{c}{350} \\
\hline\hline
\multicolumn{1}{l}{Process}	&\multicolumn{5}{c}{Rate [$10^{-5}/\mu\,{\rm m}$]}\\
\cline{2-6} ($n$,$p$)
	&1.0			&10.2	&26.9	&31.2	&32.2 \\
($\mu^-$,$\nu_\mu$)
	&12.6			&2.3		&0.9		&0.8		&0.8 \\
\hline
\multicolumn{1}{l}{Process}	&\multicolumn{5}{c}{Rate [events/kton\,day]}\\
\cline{2-6} ($n$,$p$)
	&1.4\mbox{$\times10^5$}		&1480	&61.3	&8.9		&0.1 \\
($\mu^-$,$\nu_\mu$)
	&2.1\mbox{$\times10^6$}		&390	&2.3		&0.3		&0.003 \\
\hline
Total
	&2.2\mbox{$\times10^6$} 	&1870	&63.6	&9.2		&0.1 \\
\hline
Measured
	&		&		&60		&		& \\
\end{tabular}
\end{ruledtabular}
\end{table}

In Table~\ref{tab:12b-rates}, we summarize the $^{12}$B production
rates at different depths.  For the four underground depths, the
($n$,$p$) results were obtained by direct calculations with
\textsc{fluka}, as described.  At sea level, the result was estimated
by scaling the neutron production cross section as $\rm \sigma \propto
E^{\alpha}$~\cite{olga}.  The value chosen for the $\alpha$ is the
average of the measured values on a number of unstable isotopes
produced on $^{12}$C in the beam experiment at CERN: $\alpha =
0.73$~\cite{hagner}.  The depth dependence of the $(\mu^-,\nu_\mu)$
results was obtained using the stopping muon fractions given by
Ref.~\cite{cassiday}.  Also in Table~\ref{tab:12b-rates}, the rates
per volume $R_V$ were obtained by
\begin{equation}
R_V = R_T \, \Phi_\mu \, (M_D/\rho) \, \beta\,,
\end{equation}
where $R_T$ is the rate per muon track length, $\Phi_\mu$ is the muon
flux, $M_D$ the detector mass, $\rho$ the mass density, and $\beta$ a
correction factor to compensate for averaging over the muon
spectrum~\cite{hagner}:
\begin{equation}
\beta = \frac{\langle E_\mu^\alpha \rangle}{\langle E_\mu
\rangle^\alpha} = 0.87 \pm 0.03\,.
\end{equation}
Because of how we have defined our inputs, the factor $\beta$ is only
needed for the $(n,p)$ calculations.

From the values listed in Table~\ref{tab:12b-rates}, one can see that
the dominant process for the production of $^{12}$B at depths greater
than a few hundred meters is the ($n$,$p$) exchange reaction, the
$(\mu^-,\nu_\mu)$ reaction becoming much less important.  The
systematic error on the production rate quoted in
Table~\ref{tab:12b-rates} due the uncertainty on the ($n$,$p$) cross
sections is estimated to be about 5\%.  Other uncertainties and
approximations probably increase this, but nevertheless, the
calculation is in excellent agreement with the rate measured at
2700\,m.w.e.~depth in KamLAND~\cite{kamland2}.  This agreement is an
important confirmation of entire procedure for calculating both the
secondaries produced by muons, as well as the interactions of those
secondaries.

The sea-level calculations should be taken only as crude estimates,
since one would have to properly take into account the shielding of
the detectors, non-vertical muons, unattenuated hadronic cosmic rays,
etc.  Additionally, most detectors on the surface are very small, so
that the showers induced by muons would not be fully contained;
conversely, the small size means little shielding from interactions
outside the detector.  As an example of the importance of the $^{12}$B
production rates and mechanisms, we note that the proposal of
Ref.~\cite{conrad} to measure reactor $\bar{\nu}_e + e^- \rightarrow
\bar{\nu}_e + e^-$ scattering as a test of $\sin^2{\theta_W}$; the
signal is a single scattered electron, and there is a background from
$^{12}$B beta decays~\cite{conrad}.


\section{$^{16}$N Production in Water}
\label{sec:16n}

\begin{table}[t]
\caption{\label{tab:16n-rates} Production rates for $^{16}$N in muon
induced showers at different depths $D$, given both per muon track
length, and per volume and time.}
\begin{ruledtabular}
\begin{tabular}{lddddd}
$D$ [m.w.e.]
	&\multicolumn{1}{c}{0}
	&\multicolumn{1}{c}{500}
	&\multicolumn{1}{c}{2700}
	&\multicolumn{1}{c}{3800}
	&\multicolumn{1}{c}{6000} \\
$\langle E_\mu \rangle$ [GeV]
	&\multicolumn{1}{c}{4}
	&\multicolumn{1}{c}{100}
	&\multicolumn{1}{c}{285}
	&\multicolumn{1}{c}{320}
	&\multicolumn{1}{c}{350} \\
\hline\hline
\multicolumn{1}{c}{Process}	&\multicolumn{5}{c}{Rate [$10^{-5}/\mu\,{\rm m}$]}\\
\cline{2-6} ($n$,$p$)
	&0.8			&9.1		&23.0	&25.6	&26.3 \\
($\mu^-$,$\nu_\mu$)
	&19.7			&3.6		&1.4		&1.3		&1.3 \\
\hline
\multicolumn{1}{l}{Process} &\multicolumn{5}{c}{Rate [events/kton\,day]}\\
\cline{2-6} ($n$,$p$) 
	&1.1\mbox{$\times10^5$} 	&1320	&52.4	&7.3		&0.07 \\
($\mu^-$,$\nu_\mu$)
	&2.8\mbox{$\times10^6$} 	&530	&3.2		&0.4		&0.004 \\
\hline
Total
	&2.9\mbox{$\times10^6$} 	&1850	&55.5	&7.7		&0.08 \\
\end{tabular}
\end{ruledtabular}
\end{table}

Following the same procedure as above, we also calculated the
production rates of $^{16}$N in a water-based detector.  The two
production channels taken into consideration are $^{16}$O$(n,p)^{16}$N
and $^{16}$O$(\mu^-,\nu_\mu)^{16}$N.  The cross section data for
$^{16}$O$(n,p)^{16}$N are shown in Fig.~\ref{fig:cross_sections}.
Note that by comparing the cross sections on $^{12}$C and $^{16}$O, a
somewhat lower range of neutron energies is relevant in the latter
case.  The fraction of stopping negative muons undergoing capture on
$^{16}$O in water 18.4\%~\cite{suzuki}, and the fraction of these
ending in the ground state of $^{16}$N is 10.7\%~\cite{measday}.
Results for the production rates of $^{16}$N are given in
Table~\ref{tab:16n-rates}.


\section{Concluding remarks}
\label{sec:conclusions}

\begin{figure}[t]
\includegraphics[width=3.25in]{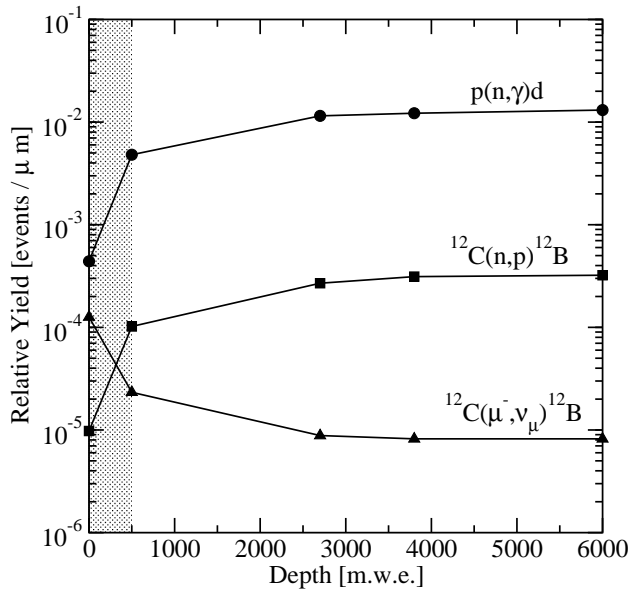}
\caption{\label{fig:yields_depth} Yields for neutron capture,
$^{12}$C$(n,p)^{12}$B, and $^{12}$C$(\mu^-,\nu_\mu)^{12}$B, as a
function of depth, in units of per muon track length.  In the
shaded region below 500 m.w.e., the values and their variation with
depth should be taken only as crude estimates.}
\end{figure}

In this paper we present a study of the production mechanism of the
$^{12}$B isotope in oil-based (organic liquid scintillator) detectors,
and also of the $^{16}$N isotope in water-based detectors.  At depths
more than a few hundred m.w.e.~underground, their production is almost
completely via $(n,p)$ reactions initiated by fast muon-induced
neutrons.  We performed an {\it ab initio} calculation of the
production rates and compared the calculated total production rate for
$^{12}$B with data measured in KamLAND, obtaining excellent agreement.
The paper offers a further validation of the technique exploited for
the calculation of the rate of production of cosmogenic isotopes,
which was previously developed in the context of the study of the
$^{11}$C rate in oil detectors~\cite{galbiati} and of several
cosmogenic isotopes in xenon~\cite{bender}.

In Fig.~\ref{fig:yields_depth} we show the variation with depth of
three reaction rates in oil-based detectors:
\begin{enumerate}

\item The $^{12}$C$(\mu^-,\nu_\mu)^{12}$B rate, a measure of the muon
flux, which is well measured and understood~\cite{cassiday}.

\item The $p(n,\gamma)d$ rate, a measure of the muon-induced
soft-neutron flux, which is reasonably well measured and
understood~\cite{boehm}.

\item The $^{12}$C$(n,p)^{12}$B rate, a measure of the muon-induced
moderate-energy neutron flux, which is uncertain~\cite{wang}; it is quite
significant that the point at 2700 m.w.e.~has been confirmed by
KamLAND~\cite{kamland2}.

\end{enumerate}
The curves for water-based detectors are similar.

Based on these results, we note that the depth dependence of these
reactions is both mild and well-understood.  Accordingly, we place the
most significance on the {\it relative heights} of the curves in
Fig.~\ref{fig:yields_depth}.  Thus in terms of further testing of the
muon-induced backgrounds underground, it is difficult to make progress by
trying to measure the mild depth dependence more precisely.  Instead,
it would likely be much more fruitful to measure isotope production
ratios at a fixed (or extrapolated) depth, since these vary by orders
of magnitude, not factors of 2.  These orders of magnitude reflect
both the strong variation of the secondary spectra with energy, as
well as the energy dependence of the associated isotope production
reactions.  For example, a measurement of the $^{12}$B production
rate, especially relative to the total neutron capture rate, directly probes
the 10--100 MeV neutron flux.

Thus, it would be very valuable if the KamLAND, Super-Kamiokande, and the
Sudbury Neutrino Observatory experiments were to publish their
detailed results on the relative yields of unstable isotopes produced by
muons, as a function of distance from the muon track.  It would be
especially useful to have results on the correlations in particle yields, i.e.,
which isotopes (including neutrons) accompany each other in a given
spallation interaction, and at what distances.  In the Sudbury Neutrino
Observatory, absolute muon rates are very low, which restricts the
possible statistics; however, since their intrinsic and muon
background rates are so low, there is a unique opportunity to measure
all detector activity following a muon out to very large distances and
times.

The development of a well-tested physical model for all secondaries
induced by muons would very likely allow more precise cuts in existing
experiments, some of which have $\sim 20\%$ deadtime due to cuts
following muons.  It would also lead to better design considerations
for future experiments pursuing greater sensitivity for reactor
neutrinos~\cite{reactor}, low-energy solar neutrinos~\cite{lowesolar},
the diffuse supernova neutrino background~\cite{dsnb},
double beta decay\cite{betabeta}, and dark matter~\cite{darkmatter}.


\vspace{-0.25cm}
\acknowledgments
\vspace{-0.25cm}
We thank A.~Ianni, V. Kudryavtsev, J. Orrell, A.~Pocar, and P. Vogel
for reading the manuscript and for useful comments.
The work of C.G. was supported in part by the U.S. National Science Foundation 
under grant \mbox{PHY-0201141}.  The work of J.F.B. was supported by The 
Ohio State University.


\end{document}